\documentclass[prl,onecolumn,amsmath,amssymb]{revtex4}
\usepackage{amsmath, amssymb, graphics, graphicx}
\usepackage{textcomp, subfigure}
\usepackage{color}
\usepackage[final]{pdfpages}

\definecolor{linkcolor}{rgb}{0,0,0.6} 
\usepackage[pdftex,colorlinks=true, pdfstartview=FitV, linkcolor= linkcolor, citecolor= linkcolor, urlcolor= linkcolor, hyperindex=true,hyperfigures=true]{hyperref} 

\newcommand{\average}[1]{\left<{#1}\right>}

\newcommand{\pq}[1]{\left[{#1}\right]}

\newcommand{\albb}[1]{\textcolor{blue}{#1}}
\newcommand{\smeq}{\sigma_{m,\mathrm{eq}}}
\newcommand{\sueq}{\sigma_{1,\mathrm{eq}}}

\begin{document}
\title{\bf On the heat flux and entropy produced by  thermal fluctuations}
\author{ S. Ciliberto$^1$,  A.Imparato$^2$, A. Naert$^1$, M. Tanase $^1$}
\affiliation{1 Laboratoire de Physique,  \'Ecole Normale Sup\'erieure, C.N.R.S. UMR5672 \\ 46 All\'ee d'Italie, 69364 Lyon, France \\}
\affiliation{2 Department of Physics and Astronomy, University of Aarhus\\ Ny Munkegade, Building 1520, DK--8000 Aarhus C, Denmark}

\begin{abstract}
{
We report  an experimental and theoretical analysis of the energy exchanged between two conductors  kept at different temperature and coupled by the electric thermal noise. Experimentally we determine, as functions of the temperature difference, the heat flux, the out-of- equilibrium variance   and a conservation law for the fluctuating entropy, which we justify theoretically. The system is ruled by the same  equations as two Brownian particles kept at different temperatures and  coupled by an elastic force. Our results  set strong constrains on the energy exchanged between coupled nano-systems held at different temperatures.}
\end{abstract}
\maketitle

The fluctuations of   thermodynamics variables play an important role in understanding the  out-of-equilibrium  dynamics of small systems \cite{sek10,Seifert_2012}, such as Brownian particles~\cite{bli06,Jop,Ruben,Evans02,seifbech2012}, molecular motors~\cite{kumiko} and  other small devices \cite{Ciliberto}.    
 The  statistical properties of  work, heat and entropy,
 have been analyzed, within the context of the fluctuation theorem~\cite{gallavotti}  and stochastic thermodynamics \cite{sek10,Seifert_2012}, in several 
 experiments on  systems in contact with a single heat bath and driven out-of-equilibrium by external forces or fields \cite{bli06,Jop,Ruben,Evans02,kumiko,Ciliberto,seifbech2012}. In contrast, the important case in which the system is driven out-of-equilibrium by a temperature difference and  energy
exchange is  produced only by the thermal noise has  been analyzed only theoretically on model systems \cite{Deridda,Jarz2004,VandenBroeck,Villamania,Visco,Dhar2007,Gas2009,evans_temp,Hanggi2011} but never in an experiment because of the intrinsic difficulties of dealing with large temperature differences in small systems.  

\begin{figure}[h]
     \centering
          \includegraphics[width=0.235\textwidth]{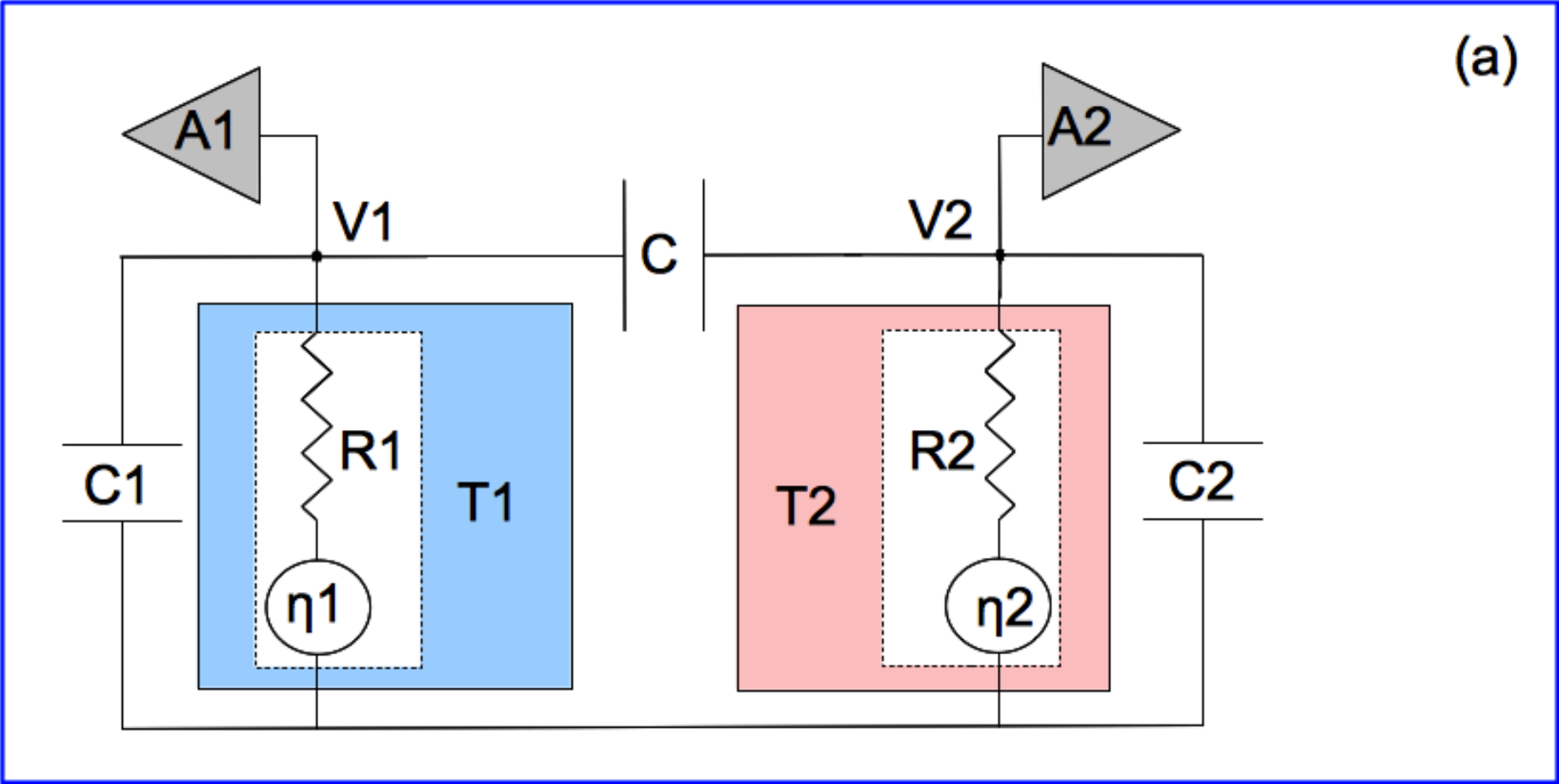}
          \includegraphics[width=0.235\textwidth]{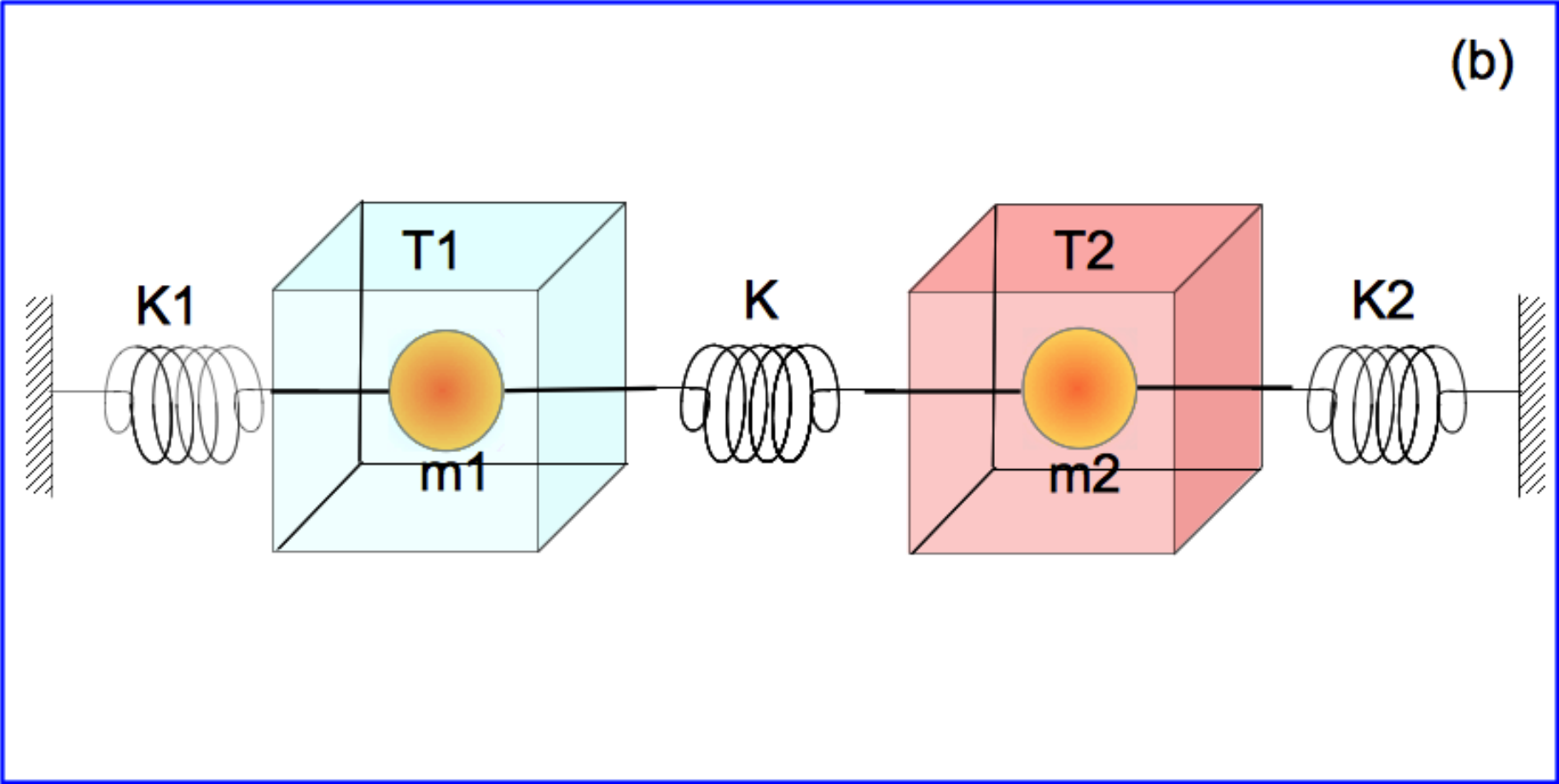}
     \caption{ a) Diagram of the circuit. The resistances $R_1$ and $R_2$ are kept at temperature $T_1$ and $T_2=296K$ respectively. They are coupled via the capacitance $C$. The capacitances $C_1$ and $C_2$ schematize the capacitance of the cables and of the amplifier inputs. 
The voltages  $V_1$ and $V_2$ are amplified  by the two low noise amplifiers $A_1$ and $A_2$ \cite{RSI}.
 b) The circuit  in  a) is  equivalent to two Brownian particles ($m_1$ and $m_2$) moving inside two different heat baths at $T_1$ and $T_2$. The two particles are trapped by two elastic potentials of stiffness $K_1$ and $K_2$ and coupled by a spring of stiffness $K$ (see text and eqs.\ref{lan1},\ref{lan2})
The analogy with the Feymann ratchet can be made by assuming as done in ref.\cite{VandenBroeck} that the particle $m1$ has an asymmetric shape and on average moves faster in one direction than in the other one.
}
     \label{fig:circuit}
\end{figure}

We report here an experimental and theoretical analysis of the statistical properties of  the energy exchanged  between two conductors  kept at different temperature and coupled by the electric thermal noise, {as depicted in fig.~\ref{fig:circuit}a}.  This system is 
 inspired by the proof developed by  Nyquist \cite{Nyquist} in order to give a theoretical explanation of the measurements of Johnson~\cite{Johnson} on the thermal  noise voltage  in conductors. In his proof, assuming thermal equilibrium between the two conductors, he deduces 
 the Nyquist noise spectral density.  At that time, well before Fluctuation Dissipation Theorem (FDT), this was the second example,  after the Einstein relation for Brownian motion, relating  the  dissipation of a system to the amplitude of the thermal noise. In this letter  we analyze the consequences  of removing the Nyquist's equilibrium conditions and we study the statistical properties of  the energy exchanged between  the  two conductors kept at different temperature.
This system  is probably among the simplest examples where  recent ideas of stochastic thermodynamics can be tested but in spite of its simplicity the explanation of the observations is far from trivial. 
{We measure experimentally  the heat flowing between the two heath baths, and show that the fluctuating entropy exhibits a conservation law.}  This system is very general because 
 is ruled by the same  equations of two Brownian particles kept at different temperatures and  coupled by an elastic force \cite{VandenBroeck,Villamania}.
Thus it gives more insight into the properties of the heat flux produced by mechanical coupling,  in the famous Feymann ratchet \cite{Feymann,Smoluchowski,Abb2000} widely studied theoretically \cite{VandenBroeck} but never in an experiment. 
Therefore our results have implications well beyond the simple system we consider here. 

Such a system is sketched in fig.\ref{fig:circuit}a). It is constituted by two resistances $R_1$ and $R_2$, which are kept at different temperature $T_1$ and $T_2$ respectively. These temperatures are controlled  by thermal baths and $T_2$ is kept fixed at $296K$ whereas $T_1$  can be set at a value  between $296K$ and $88K$ using liquid nitrogen vapor as a circulating coolant.  
In the figure, the two resistances have been drawn with their associated thermal noise generators $\eta_1$ and $\eta_2$, whose power spectral densities  are given by the Nyquist formula $|\tilde \eta_m|^2= 4 k_B R_mT_m$, with $m=1,2$ (see  eqs.\ref{lan1},\ref{lan2} and ref.\cite{Supplementary}).
   The coupling capacitance $C$ controls the electrical power exchanged between the resistances  and as a consequence the energy exchanged between the two baths. {No other coupling exists between the two resistances which are inside two separated screened boxes}. 
The quantities $C_1$ and $C_2$ are the capacitances of the circuits and the cables. 
Two extremely  low noise amplifiers  $A_1$ and $A_2$ \cite{RSI} measure the voltage $V_1$ and $V_2$ across  the resistances $R_1$ and $R_2$ respectively. {All the relevant quantities considered in this paper can be derived by the measurements of $V_1$ and $V_2$, as discussed below}. 
In the following we will take  $C=100pF,  C_1=680pF, C_2=420pF$ and $R_1=R_2=10M\Omega$, if not differently stated. 
When $T_1=T_2$ the system is in equilibrium and exhibits no net energy flux  between the two reservoirs. 
This is indeed the condition imposed  by Nyquist to prove his formula, and we use it  {to check all the values of the circuit parameters.} 
{Applying the Fluctuation-Dissipation-Theorem (FDT) to the circuit, one finds the  Nyquist's expression for the variance  of $V_1$ and $V_2$ at equilibrium, which reads  $ \smeq^2(T_m)= {k_B T_m (C+C_m')/X }$ with $X=C_2\, C_1+C\, (C_1 \, +C_2) $,  $m'=2$ if $m=1$ and $m'=1$ if $m=2$.}  For example one can check that at $T_1=T_2=296$ K, using  the above mentioned values of the capacitances and resistances,  the predicted equilibrium standard deviations of  $V_1$ and $V_2$ are  $2.33 \mu V$ and $8.16 \mu V$ respectively. These are indeed the measured values  with an accuracy better than $ 1 \% $, {see  ref.~\cite{Supplementary}  for further details on the system calibration}.
 
\begin{figure}[h]
     \centering
       \includegraphics[width=0.22\textwidth]{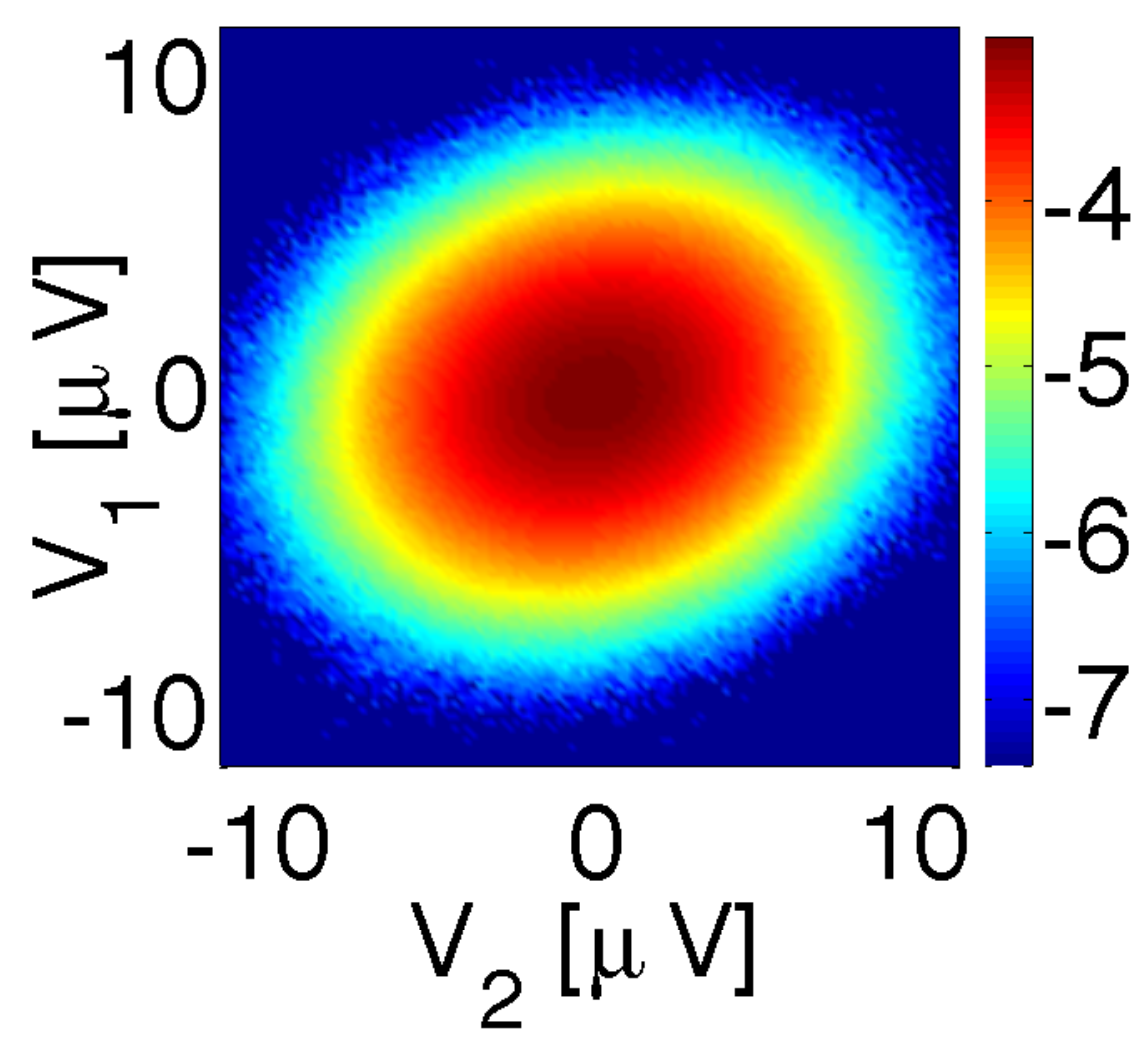}
    \includegraphics[width=0.22\textwidth]{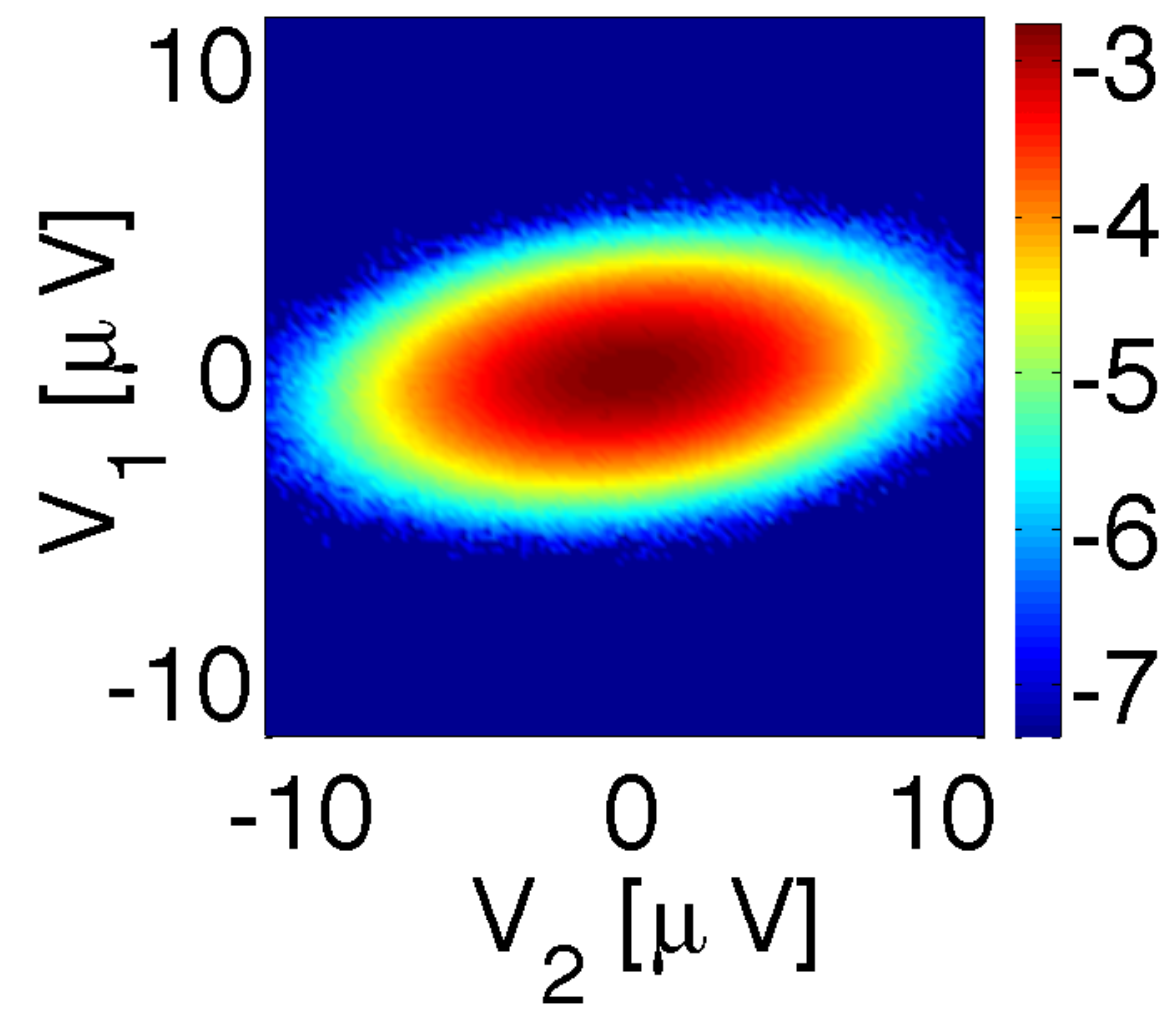}
     \caption{The joint probability $\log_{10}P(V_1,V_2)$  measured at $T_1=296K$ equilibrium (a) and out of equilibrium $T_1=88K$(b).
     The color scale   is indicated on the colorbar on the right side. 
      }
     \label{fig:pdfV1V2}
\end{figure}
The important quantity to consider here is the joint probability $P(V_1,V_2)$, which is plotted in fig.~\ref{fig:pdfV1V2}a) at $T_1=T_2$ and at 
fig.~\ref{fig:pdfV1V2}b) at  $T_1=88K$. The fact that the axis of the  ellipses defining the contours lines of  $P(V_1,V_2)$ are inclined with respect to the $x$ and $y$ axis indicates that there is a certain correlation between $V_1$ and $V_2$.  This correlation, produced by the electric coupling, plays a  major role in determining the mean  heat flux between the two reservoirs, as we discuss below. 
{
The interesting new features occur of course  when $T_1\ne T_2$. The questions that we address for such a system are:  What are the heat flux and the entropy production rate ? How the variance of $V_1$ an $V_2$ are modified because of the heat flux ?  What is the role of correlation between $V_1$ and $V_2$? We will see that these questions are quite relevant and have no  obvious answers because of the statistical nature of the energy transfer. 
}
\begin{figure}[h]
     \centering
    \includegraphics[width=0.23\textwidth]{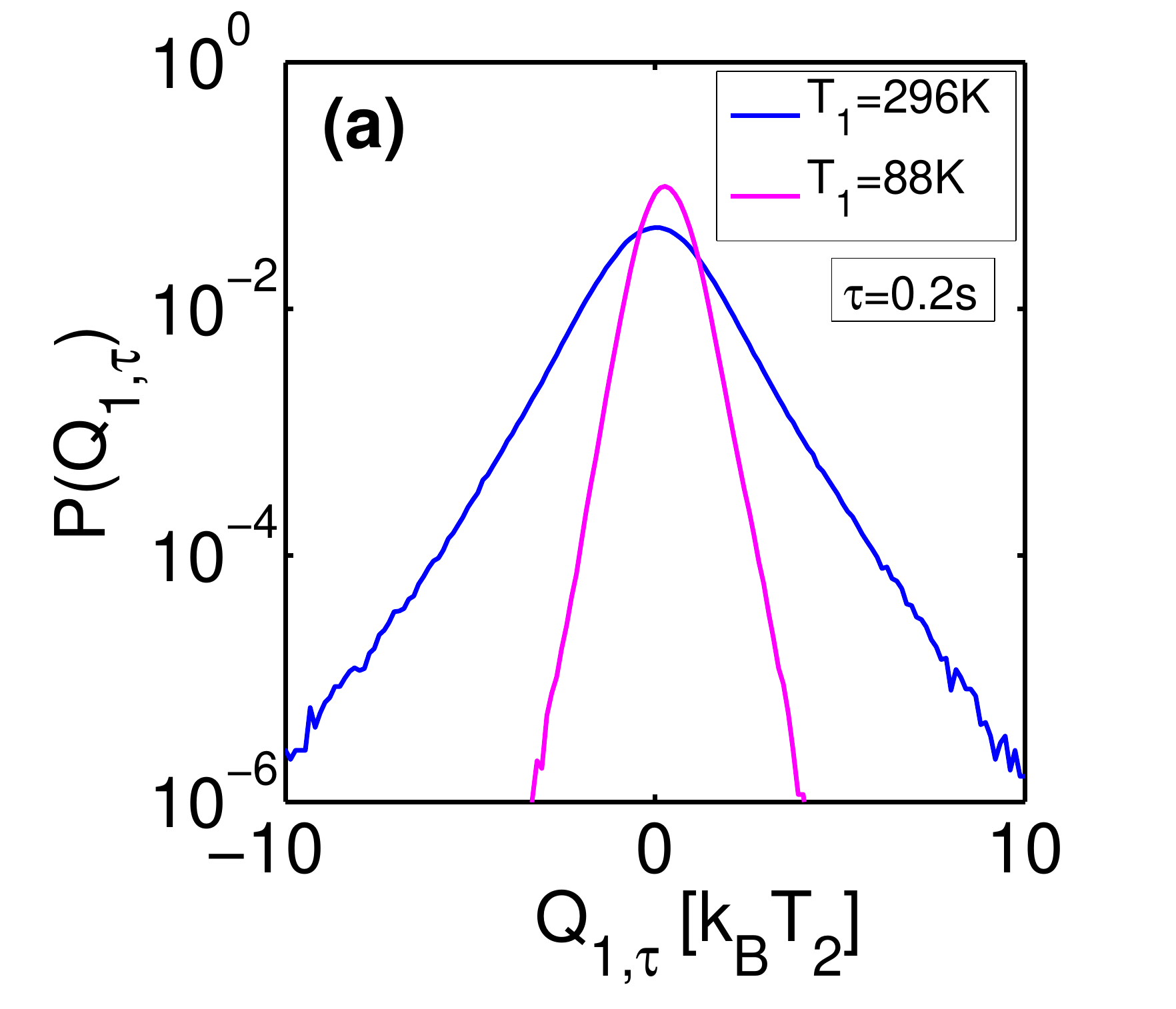} 
    \includegraphics[width=0.23\textwidth]{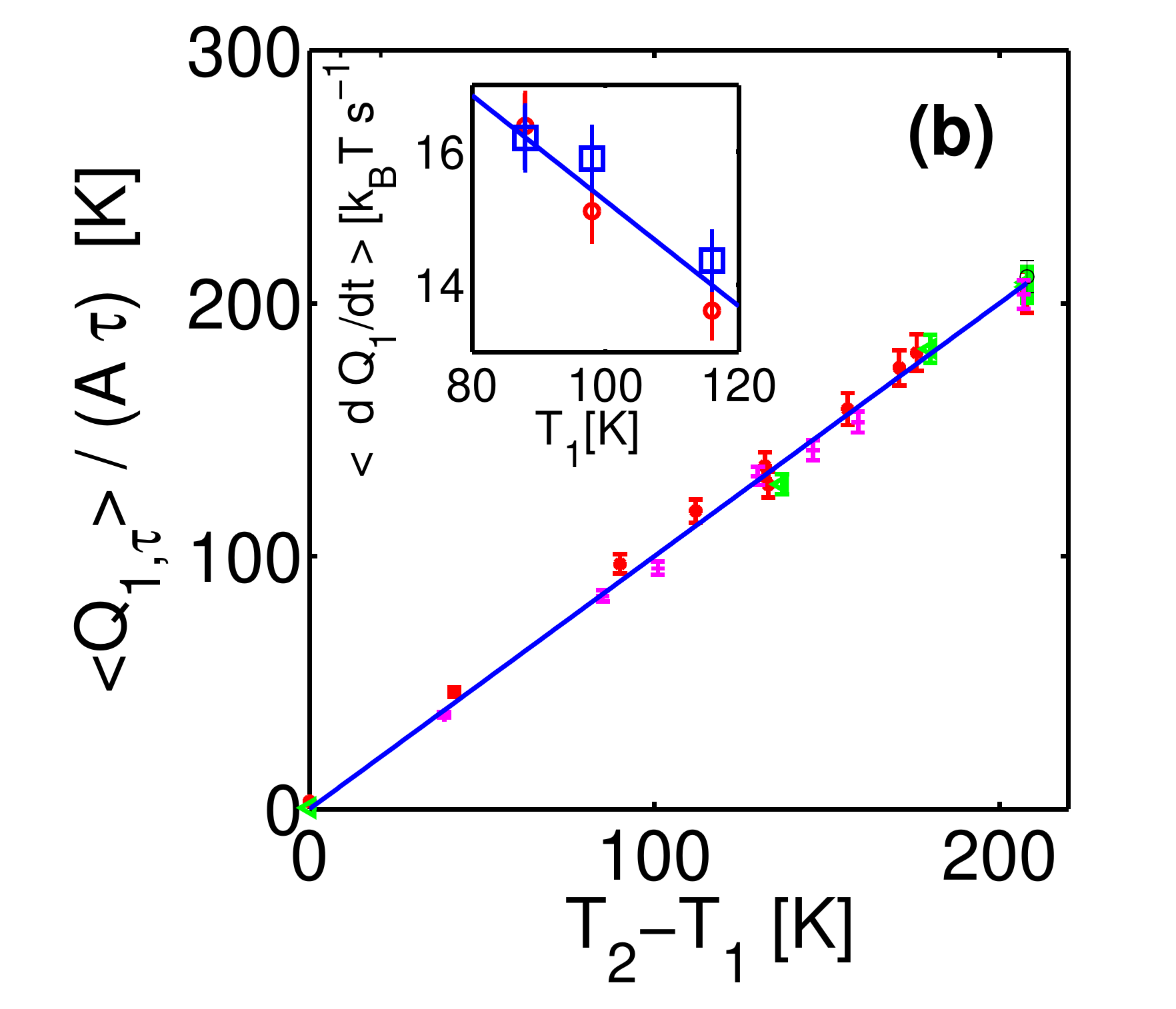}
     \caption{a) The  probability $P(Q_{1,\tau})$  measured at $T_1=296K$ (blue line) equilibrium 
    and  $T_1=88K$ (magenta line) out of equilibrium. Notice that the peak of the $P(Q_{1,\tau})$ is centered at zero at equilibrium and shifted towards a positive value out of equilibrium. The amount of the shift is very small and  {is $\sim k_B(T_2-T_1)$.}   b) The measured mean value of $\average{Q_{1,\tau}}$ is a linear  function of $(T_2-T_1)$.  The red points correspond to measurements performed with the values of the capacitance  $C_1,C_2,C$ given in the text and $\tau=0.2s$.  The other  symbols and colors pertain to different values of these capacitance and other  $\tau$: {(black $\circ$) $\tau=0.4s,C=1000pF$, (green $\triangleleft$) $\tau=0.1s,C=100pF$, (magenta $+$) $\tau=0.5s,C=100pF$}.  The values of $\average{Q_{1,\tau}}$ have been rescaled by the parameter dependent theoretical prefactor $A$, which allows the comparison of different experimental configurations. The continuous blue line with slope $1$ is the theoretical prediction of eq.~\ref{q:rate}. 
    In the inset the values of $<\dot Q_1>$ (at $C=1000pF$) directly measured using $P(Q_1)$ (blue square) are compared with those (red circles)  obtained  from the equality 
   $<\dot Q_1>=(\sigma_1^2-\sueq^2)/R_1$, as discussed in the text. 
}
\label{fig:P_Q}
\end{figure}

We consider the electric power dissipated in the resistance $R_m$ with  $m=1,2$ which reads  $\dot Q_m=V_m \, i_m$  where $i_m$ is the current flowing in the resistance $m$. The integral of the power  over a time $\tau$   is the total energy $Q_m$, dissipated by the resistance in this time interval, i.e. $Q_{m,\tau}= \int_t^{t+\tau} i_{m} \, V_m \,  dt $. All the voltages $V_m$ and currents $i_m$  can be  measured: indeed  we have
 $i_m=i_C-i_{C_m}$ where  $i_C=C {d (V_2-V_1) \over dt}$ is the current  flowing in the capacitance $C$, and 
$ i_{C_m}=C_m {dV_m \over dt}$ is the current flowing in $C_m$. Thus rearranging the terms one finds that 
$Q_{m,\tau}= W_{m,\tau}-\Delta U_{m,\tau}$ where $W_{1,\tau}= \int_t^{t+\tau}   C \, V_1 {d V_2 \over dt} dt$, 
 $W_{2,\tau}= \int_t^{t+\tau}  C V_2 {d V_1\over dt} dt$  and  $\Delta U_{m,\tau} = {(C_m+C)\over 2}(V_m(t+\tau)^2-V_m(t)^2)$ is the potential energy change of the circuit $m$ in the time $\tau$. Notice that $W_m$ are the terms responsible for the energy exchange since they couple the fluctuations of the two circuits. The quantities  $W_{1,\tau}$ and $W_{2,\tau}$ can  be  identified as the work performed by the circuit 2 on 1 and vice-versa \cite{Sekimoto,VanZonCil,Garnier}, respectively. Thus, the quantity $Q_{1,\tau}$ ($Q_{2,\tau}$) can  be interpreted as the heat flowing from the reservoir 2 to the reservoir 1 (from 1 to 2), in the time interval $\tau$, as an effect of the temperature difference.
As the two variables $V_m$ are fluctuating voltages all the other quantities also fluctuate. 
In fig.~\ref{fig:P_Q}a) we show the probability density function  $P(Q_{1,\tau})$, at various temperatures:
we see that $Q_{1,\tau}$ is a strongly fluctuating quantity, whose $P(Q_{1,\tau})$ has long exponential tails. 

Notice that although for $T_1< T_2$ the mean value of $Q_{1,\tau}$ is positive, instantaneous negative fluctuations can occur,  i.e., sometimes the heat flux is reversed.   The mean values of the dissipated heats are expected to be  linear functions of  the temperature difference $\Delta T=T_2-T_1$, i.e. $\average{Q_{1,\tau}}=A \,  \tau \,\Delta T$, where $A$ is a parameter dependent quantity, that can be obtained explicitly from eqs.~\ref{lan1} and \ref{lan2} below.
This relation is confirmed by our experimental results, as shown in fig.~\ref{fig:P_Q}b. Furthermore, the mean values of the dissipated heat satisfy the equality $\average{Q_2}=-\average{Q_1} $, corresponding to an energy conservation principle: the power extracted from the bath 2 is dissipated into the bath 1 because of the electric coupling.
 This mean flow produces a change of the variances 
$\sigma_m^2(T_m)$ of  
$V_m$  with respect to the equilibrium value  $\smeq^2(T_m)$, {that is the equilibrium value measured when the two baths are at the same temperature  $T_m$}.
Specifically we find    $\sigma_m^2(T_m)= \smeq^2(T_m)+ <~\dot Q_m~> R_m$  which is an extension to two temperatures of the Harada-Sasa relation \cite{harada} {(see also ref.\cite{Supplementary} for a theoretical proof of this experimental result)}. This result is shown in the inset of fig.~\ref{fig:P_Q}b) where the values of $\average{\dot Q_m}$ directly estimated from the experimental data (using the steady state $P(Q_m)$) are compared with those  obtained from the difference of the variances of  $V_1$ measured in equilibrium and out-of-equilibrium. The values are comparable within error bars and show that the out-of-equilibrium variances  are modified only by the heat flux.  
\begin{figure}[h]
     \centering
    \includegraphics[width=0.238\textwidth]{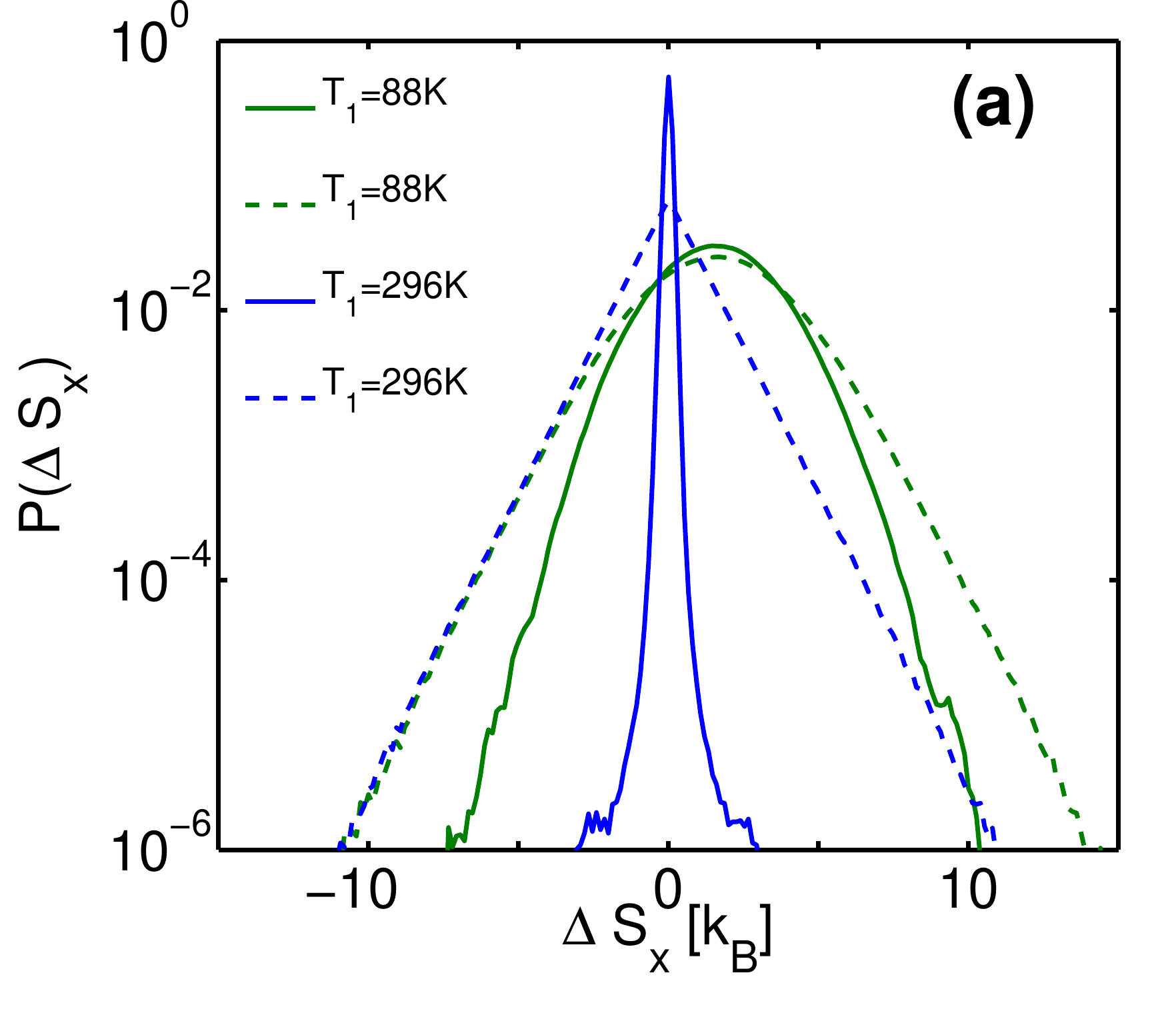}
    \includegraphics[width=0.238\textwidth]{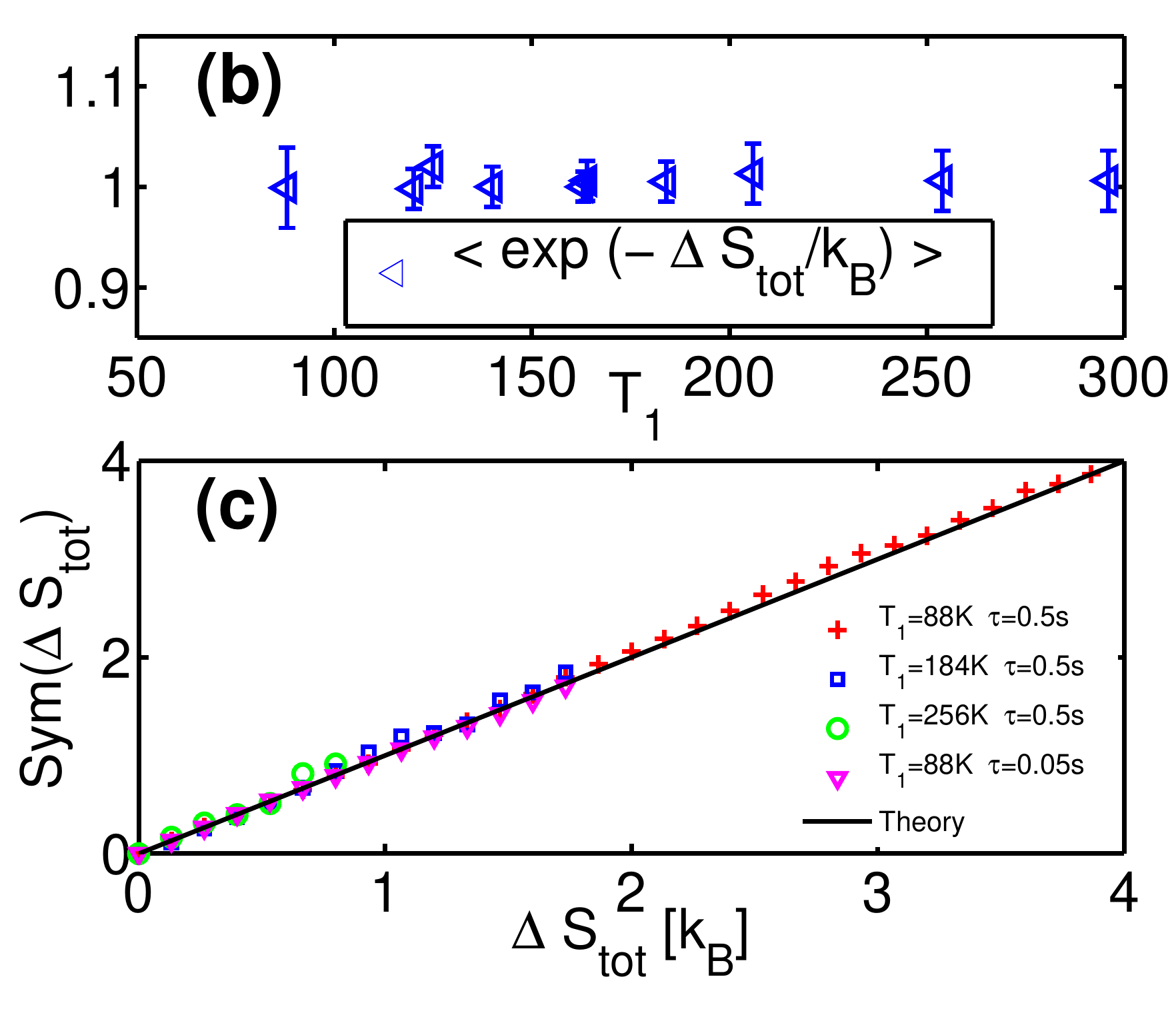}
     \caption{ a) The  probability $P(\Delta S_r)$ (dashed lines) and $P(\Delta S_{tot})$ (continuous lines)   measured at $T_1=296K$ (blue line) which corresponds to equilibrium 
    and  $T_1=88K$ (green lines) out of equilibrium. Notice that both distributions are  centered at zero at equilibrium and shifted towards positive value in the out-of-equilibrium.    b)  $\average{\exp(-\Delta S_{tot})}$ as a function of $T_1$ at two different $\tau=0.5s$ and $\tau=0.1s$.  c) Symmetry function $\rm{Sym}(\Delta S_{tot})=\log[P(\Delta S_{tot})/P(-\Delta S_{tot})]$ as a function of $\Delta S_{tot}$. The black straight line of slope 1 corresponds to the theoretical prediction. }
     \label{fig:P_DS}
\end{figure}
It is  now important to analyze the entropy produced by the  total system, circuit plus heat reservoirs.  We consider first the entropy  $\Delta S_{r,\tau}$ due to the heat exchanged with the reservoirs, which reads  $\Delta S_{r,\tau}= Q_{1,\tau}/T_1 +Q_{2,\tau}/T_2$. This entropy is a fluctuating quantity as both $Q_1$ and  $Q_2$ fluctuate, and its  average in a time $\tau$ is 
$\average{\Delta S_{r,\tau}} = \average{Q_{r,\tau}}(1/T_1-1/T_2)=A \tau (T_2-T_1)^2 /(T_2 \, T_1)$.   However the reservoir entropy $\Delta S_{r,\tau}$ is not the only component of the total entropy production: one has to take into account the  entropy variation of the system, due to its dynamical evolution.
Indeed, the state variables $V_m$  also fluctuate as an effect of the thermal noise, and thus, if one measures their values at regular time interval, one obtains a ``trajectory'' in the phase space $(V_1(t), V_2(t))$. Thus, following Seifert \cite{Seifert}, who developed this concept for a single heat bath, one can introduce a trajectory  entropy for the evolving system $S_s(t)=-k_B \log P (V_1(t), V_2(t))$, which extends to non-equilibrium systems the standard Gibbs entropy concept.  Therefore, when evaluating the total entropy production, one has to take into account the contribution over the time interval $\tau$ of 
\begin{equation}
\Delta S_{s,\tau}= - k_B  \log \left[{ P(V_1(t+\tau ),V_2(t+\tau )) \over P(V_1(t),V_2(t))} \right].
\label{eq:DS_tot}
\end{equation}
It is worth noting that the system we consider is in a non-equilibrium steady state, with a constant external driving $\Delta T$. Therefore the probability distribution $P(V_1,V_2)$ (as shown in fig.~\ref{fig:pdfV1V2}b)) does not depend explicitly on the time, and   $\Delta S_{s,\tau}$
is non vanishing whenever the final point of the trajectory is different
from the initial one: $(V_1(t+\tau ),V_2(t+\tau ))\neq (V_1(t),V_2(t))$.
Thus the total entropy {change} reads $\Delta S_{tot,\tau}=\Delta  S_{r,\tau} +\Delta S_{s,\tau}$, where we omit the explicit dependence on $t$, as the system is in a steady-state as discussed above. This entropy has  several interesting features. The first  one is  that $\average{\Delta S_{s,\tau}}=0 $, and as a consequence $\average{\Delta S_{tot}}=\average{ \Delta S_r}$  which grows with increasing  $\Delta T$. The second and
 most interesting result is that independently of $\Delta T$ and of $\tau$, the following equality always holds: 
\begin{equation}
\average{\exp(-\Delta S_{tot} /k_B)}=1,
\label{eq:DS}
\end{equation}
for which we find both experimental evidence, as discussed in the following, and provide a theoretical proof in  ref.~\cite{Supplementary}. Equation~(\ref{eq:DS}) represents an extension to two temperature sources of the result obtained for  a system in a single heat bath driven out-of-equilibrium by a time dependent mechanical force~\cite{Seifert,Evans02} and our results provide the first experimental verification of the expression in  a system driven by a temperature difference. Eq.~(\ref{eq:DS}) implies that $\average{\Delta S_{tot} } \, \ge  0$, as prescribed by the second law.  {From symmetry considerations, it follows immediately that, at equilibrium ($T_1=T_2$), the probability distribution of $\Delta S_{tot}$ is symmetric: $P_{eq}(\Delta S_{tot})=P_{eq}(-\Delta S_{tot})$. Thus Eq.~(\ref{eq:DS}) implies  that  the probability density function of $\Delta S_{tot}$ is a Dirac $\delta$ function  when $T_1=T_2$, i.e. the quantity $\Delta S_{tot}$ is rigorously zero in equilibrium, both in average and fluctuations, and so its mean value  and  variance provide a measure of the entropy production.}
{The  measured probabilities $P(\Delta S_r)$ and $P(\Delta S_{tot})$ are  shown in fig. \ref{fig:P_DS}a). We see that  $P(\Delta S_{r})$  and
 $P(\Delta S_{tot})$ are quite different and that the latter is close to a Gaussian and reduces to a Dirac $\delta$ function in equilibrium, i.e. $T_1=T_2=296K$ (notice that, in fig.\ref{fig:P_DS}a,  the small broadening of the equilibrium $P(\Delta S_{tot})$ is just due to unavoidable experimental noise and  discretization of the experimental probability density functions). }The experimental measurements satisfy eq.~(\ref{eq:DS}) as it is shown in  fig. \ref{fig:P_DS}b). It is worth to note that eq.~(\ref{eq:DS}) implies that  $P(\Delta S_{tot})$ should satisfy a fluctuation theorem of the form
    $\log [P(\Delta S_{tot})/P(-\Delta S_{tot})]= \Delta S_{tot}/k_B, \, \, \,  \forall \tau,\Delta T $, as discussed extensively in reference \cite{EVDB10,Seifert_2012}.  We clearly see in fig.\ref{fig:P_DS}c) that this relation holds for different values of the temperature gradient.    
Thus this experiment clearly establishes  a relationship between the mean and the variance of the entropy production rate 
in a system driven out-of-equilibrium  by the temperature difference between two thermal baths coupled by electrical noise. Because of the formal analogy with Brownian motion the results also apply to mechanical coupling as discussed in the following. 

We will now give a theoretical interpretation of the experimental observations. This will  allow us to show  the analogy of our system with two interacting  Brownian particles coupled  to two different temperatures,  see fig.~\ref{fig:circuit}-b).   
Let $q_m$ ($m=1,2$) be the charges that have flowed through the resistances $R_m$, so  the instantaneous current flowing through them is $i_m=\dot q_m $. 
 A  circuit analysis shows that the equations for the charges are: 
 \begin{eqnarray}
 R_1 \dot q_1&=&- q_1\, {C_2 \over X}+  (q_2-q_1){C \over X} + \eta_1  \label{lan1}\\
 R_2 \dot  q_2&=&- q_2\, {C_1  \over X}+  (q_1-q_2){C \over X} + \eta_2 \label{lan2}
\end{eqnarray}
where   $\eta_m$ is the usual white noise: $\average{\eta_i(t)\eta_j(t')}=2 \delta_{ij} {k_BT_i R_j} \delta(t-t')$.
The relationships between the measured voltages and the charges are: 
\begin{eqnarray}
q_1&=& (V_1-V_2) \, C + V_1\, C_1  \label{eq_q1}\\
q_2&=& (V_1-V_2) \, C - V_2\, C_2  \label{eq_q2}
\end{eqnarray}
Eqs.~\ref{lan1} and \ref{lan2} are the same of  those for the two coupled Brownian particles sketched in fig.\ref{fig:circuit}b)  by considering 
$q_m$ the displacement of the particle $m$, $i_m$ its velocity,  $K_m=1/C_m$ the stiffness of the spring $m$, $K=1/C$ the coupling spring and $R_m$ the viscosity. 
With this analogy we see that our definition of the heat flow $Q_m$ corresponds
exactly to the work performed by the viscous forces and by the bath on
the particle $m$, and it is consistent with the stochastic thermodynamics
definition  \cite{Sekimoto,VanZonCil,Seifert_2012,alb1,alb2}. 
{\hskip 1cm}
Thus our theoretical analysis and the experimental results apply to both interacting  mechanical and electrical systems coupled to baths at different temperatures. 
Starting from eqs.~(\ref{lan1})-(\ref{lan2}), we can  prove  (see ref.~\cite{Supplementary}) that  eq.\ref{eq:DS} is an exact result and that the average dissipated heat rate is 
\begin{equation}
\average{\dot Q_{1}}=A \, (T_2-T_1)=\frac{C^2 \Delta T}{X Y},
\label{q:rate}
\end{equation}
with $Y=\pq{(C_1+C) R_1 +(C_2+C) R_2 }$ and $A=C^2/(X \, Y)$ is the parameter used to rescale the data in fig.~\ref{fig:P_Q}b). 

To conclude we have studied experimentally the statistical properties of the energy exchanged between two  heat baths at different temperature which are coupled by electric thermal noise. We have measured the heat flux, the entropy production rate and we have shown the existence of a conservation law for entropy which imposes the existence of a fluctuation theorem which is not asymptotic in time.  Our results, which are theoretically proved, are very general since the electric system considered here is ruled by 
the same equations as for two Brownian particles, held at  different temperatures and mechanically coupled. Therefore  these results set precise constraints on the energy exchanged between coupled nano and micro-systems held at different temperatures. We finally mention that for the quantity $W_i$ an asymptotic fluctuation theorem can be proved both experimentally and theoretically, and this will be the subject of a paper in preparation.

\newpage


\section*{Supplementary material}

# On the heat flux and entropy produced by thermal fluctuations: Supplementary information

S. Ciliberto[1], A. Imparato[2], A. Naert[1], M. Tanase [1]

*1 Laboratoire de Physique,
École Normale Supérieure, C.N.R.S. UMR5672
46 Allée d'Italie, 69364 Lyon, France
and
2 Department of Physics and Astronomy, University of Aarhus
Ny Munkegade, Building 1520, DK–8000 Aarhus C, Denmark*

## I. EXPERIMENTAL DETAILS

### A. Experimental set up

The electric systems and amplifiers are inside a Faraday cage and mounted on a floating optical table to reduce mechanical and acoustical noise. The resistance $R_1$, which is cooled by liquid Nitrogen vapors, changes of less than $0.1\%$ in the whole temperature range. Its temperature is measured by a PT1000 which is inside the same shield of $R_1$. The signal $V_1$ and $V_2$ are amplified by two custom designed JFET amplifiers [1] with an input current of $1pA$ and a noise of $0.7nV\sqrt{Hz}$ at frequencies larger than $1Hz$ and increases at $8nV\sqrt{Hz}$ at $0.1Hz$. The resistances $R1$ and $R2$ have been used as input resistances of the amplifiers. The two signals $V_1$ and $V_2$ are amplified $10^4$ times and the amplifier outputs are filtered (at $4KHz$ to avoid aliasing) and acquired at $8KHz$ by 24 bits-ADC. We used different sets of $C_1, C_2$ and $C$. The values of $C1$ and $C2$ are essentially set by the input capacitance of the amplifiers and by the cable length $680pF < C1 < 780pF$ and $400pF < C_2 < 500pF$. Instead $C$ has been changed from $100pF$ to $1000pF$. The system has always been calibrated in equilibrium at $T_1 = T_2 = 296K$ using the FDT and estimating the spectrum using the values of the capacitances, see next sections.

### B. Noise spectrum of the amplifiers

The noise spectrum of the amplifiers $A_1$ and $A_2$ (Fig.1 of the main text), measured with a short circuit at the inputs, is plotted in fig.S.1a) and compared with the spectrum $Sp_1$ of $V_1$ at $T_1 = 88K$. We see that the useful signal is several order of magnitude larger than the amplifiers noise.

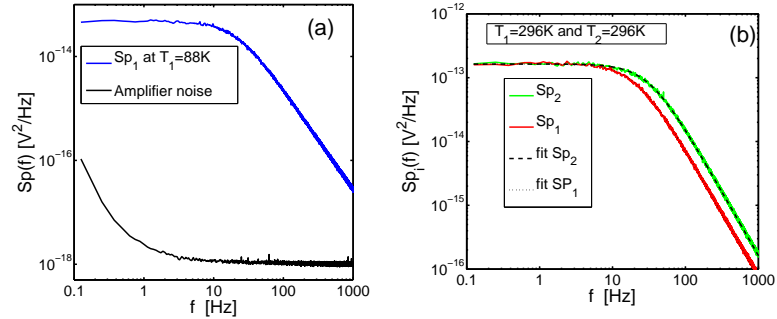

FIG. S.1: a) The power spectra $Sp_1$ of $V_1$ measured at $T_1 = 88K$ (blue line) ( $C = 100pF, C_1 = 680pF, C_2 = 430pF$) is compared to the spectrum of the amplifier noise. b) The equilibrium spectra $Sp_1$ ( red line) and $Sp_2$ (green line) measured at $T_1 = T_2 = 296K$ are compared with prediction of eqs.S.1 and S.2 in order to check the values of the capacitances ($C_1, C_2$).

## C. Check of the calibration

The equilibrium spectra of $V_1$ and $V_2$ at $T_1 = T_2$ used for calibration of the capacitances are:

$$S_1(\omega) = \frac{4k_BT_1\, R_1[1 + \omega^2(C^2R_1R_2 + R_2^2(C_2+C)^2)]}{(1 - \omega^2\, X\, R_1R_2)^2 + \omega^2Y^2} \tag{S.1}$$

$$S_2(\omega) = \frac{4k_BT_2\, R_2[1 + \omega^2(C^2R_1R_2 + R_1^2(C_1+C)^2)]}{(1 - \omega^2\, X\, R_1R_2)^2 + \omega^2Y^2} \tag{S.2}$$

where $Y = [(C_1+C)R_1 + (C_2+C)R_2]$ and $X = C_2\,C_1 + C\,(C_1+C_2)$. This spectra can be easily obtained by applying FDT to the circuit of fig.1 in the main text.

The two computed spectra are compared to the measured ones in fig. S.1a). This comparison allows us to check the values of the capacitances $C_1$ and $C_2$ which depend on the cable length. We see that the agreement between the prediction and the measured power spectra is excellent and the global error on calibration is of the order of 1%. This corresponds exactly to the case discussed by Nyquist in which the two resistances at the same temperature are exchanging energy via an electric circuit ($C$ in our case).

## D. The power spectra of $V_1$ and $V_2$ out-of-equilibrium

When $T_1 \neq T_2$ the power spectra of $V_1$ and $V_2$ are:

$$S_1(\omega) = \frac{4k_BT_1\, R_1[1 + \omega^2(C^2R_1R_2 + R_2^2(C_2+C)^2)]}{(1 - \omega^2\, X\, R_1R_2)^2 + \omega^2Y^2} + \frac{4k_B(T_2-T_1)\,\omega^2\, C^2R_1^2R_2}{(1 - \omega^2\, X\, R_1R_2)^2 + \omega^2Y^2} \tag{S.3}$$

$$S_2(\omega) = \frac{4k_BT_2\, R_2[1 + \omega^2(C^2R_1R_2 + R_1^2(C_1+C)^2)]}{(1 - \omega^2\, X\, R_1R_2)^2 + \omega^2Y^2} + \frac{4k_B(T_1-T_2)\,\omega^2\, C^2R_2^2R_1}{(1 - \omega^2\, X\, R_1R_2)^2 + \omega^2Y^2} \tag{S.4}$$

These equations have been obtained by Fourier transforming eqs. S.7,S.8, solving for $\tilde{V}_1(\omega)$ and $\tilde{V}_2(\omega)$ and computing the modula. The integral of eqs. S.3 and S.4 gives the variances eq. S.24 directly computed from the distributions.

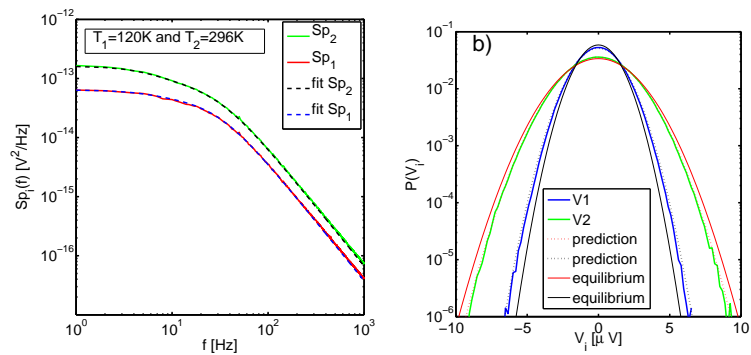

FIG. S.2: a) The power spectra $Sp_1$ of $V_1$ and $Sp_2$ of $V_2$ measured at $T_1 = 120K$ and $T_2 = 296K$ ($C = 100pF, C_1 = 680pF, C_2 = 430pF$) are compared with the prediction of eq.S.3 and S.4 (dashed lines) b) The corresponding Probability Density Function $P(V_1)$ of $V_1$ (green line) and $P(V_2)$ of $V_2$ (blue line) measured at $T_1 = 120K$ and $T_2 = 296K$. Dotted lines are the out-of-equilibrium PDF, whose variance is estimated from the measure of the heat flux (see fig.3 in the main text) and eq.S.24 in the following section. The continuous red line is the equilibrium $P(V_2)$ at $T_1 = 296K$ and the black continuous line corresponds to the equilibrium $P(V_1)$ at $T_2 = 120K$.

## E. Measure of the equilibrium variance of $V_1$ as a function of $T_1$

This measure is necessary to estimate $<Q_1>$ starting from the measurement of the variances as explained in fig.4 of the main text. We first measure $\sigma_{m,\text{eq}}^2(T1)$ at $T_1 = T_2 = 296K$. Indeed in equilibrium the variance must be proportional to

$T_1 = T_2$, i.e. $\sigma^2_{m,\text{eq}}(T_1) = \alpha_m T_1$, and from the equilibrium measurements at $T_1 = T_2$ one gets the proportionality constant $\alpha_m = \sigma^2_{m,\text{eq}}(T_1)/T_1$. Thus when $T_1 < T_2$ one can estimate the values of the equilibrium variances $\sigma^2_{m,\text{eq}}(T1) = \alpha_m T_1$. As explained in the main text $<\dot{Q}_1> = (\sigma_1^2(T1) - \sigma^2_{m,\text{eq}}(T1))/R_1$. In fig. S.2b) we compare the measured PDF of $V_1$ and $V_2$ with the equilibrium and the out-of-equilibrium distributions as computed by using the theoretical predictions eq.S.24 for the variance.

## II. DYNAMICAL EQUATIONS FOR $V_m$ AND $Q_m$.

We want to describe, with a set of coupled Langevin equations, the dynamical evolution of both the electric and thermodynamic variables for the circuit in fig. 1 of the main text. For this purpose we write the Langevin equations governing the dynamical evolutions for the voltages across the circuit:

$$(C_1 + C)\dot{V}_1 = C\dot{V}_2 + \frac{1}{R_1}(\eta_1 - V_1) \tag{S.5}$$

$$(C_2 + C)\dot{V}_2 = C\dot{V}_1 + \frac{1}{R_2}(\eta_2 - V_2) \tag{S.6}$$

where we have substituted eqs. (5)-(6) into eqs. (3)-(4) in the main text. We rearrange these equations in a standard form, and obtain

$$\dot{V}_1 = f_1(V_1, V_2) + \sigma_{11}\eta_1 + \sigma_{12}\eta_2 = f_1(V_1, V_2) + \xi_1 \tag{S.7}$$

$$\dot{V}_2 = f_2(V_1, V_2) + \sigma_{21}\eta_1 + \sigma_{22}\eta_2 = f_2(V_1, V_2) + \xi_2 \tag{S.8}$$

where the "forces" acting on the circuits read

$$f_1(V_1, V_2) = \alpha_1 V_1 + \alpha_2 V_2 = -\frac{C_2 R_2 V_1 + C(R_2 V_1 + R_1 V_2)}{[C_2 C + C_1(C_2 + C)]R_1 R_2}, \tag{S.9}$$

$$f_2(V_1, V_2) = \gamma_1 V_1 + \gamma_2 V_2 = -\frac{C_1 R_1 V_2 + C(R_2 V_1 + R_1 V_2)}{[C_2 C + C_1(C_2 + C)]R_1 R_2}, \tag{S.10}$$

the coefficients $\sigma_{ij}$ read

$$\sigma_{11} = \frac{C_2 + C}{XR1}$$

$$R_2\sigma_{12} = R_1\sigma_{21} = \frac{C}{X}$$

$$\sigma_{22} = \frac{C_1 + C}{XR_2},$$

and the noises $\xi_i$ introduced in eqs. (S.7)-(S.8) are now correlated $\langle \xi_i \xi_j' \rangle = 2\theta_{ij}\delta(t-t')$, where

$$\theta_{11} = \frac{T_1(C_2+C)^2}{R_1(C_2 C + C_1(C_2+C))^2} + \frac{T_2 C^2}{R_2(C_2 C + C_1(C_2+C))^2}, \tag{S.11}$$

$$\theta_{12} = \frac{T_1(C(C_2+C))}{R_1(C_2 C + C_1(C_2+C))^2} + \frac{T_2(C(C_1+C))}{R_2(C_2 C + C_1(C_2+C))^2}, \tag{S.12}$$

$$\theta_{22} = \frac{T_1 C^2}{R_1(C_2 C + C_1(C_2+C))^2} + \frac{T_2(C_1+C)^2}{R_2(C_2 C + C_1(C_2+C))^2}, \tag{S.13}$$

and $\theta_{12} = \theta_{21}$. We now notice that the rate of the dissipated heat in circuit $m$ reads

$$\dot{Q}_m = V_m i_m = \frac{V_m}{R_m}(V_m - \eta_m) = V_m \left[ (C_m + C)\dot{V}_m - C\dot{V}_{m'} \right], \tag{S.14}$$

where $m' = 2$ if $m = 1$, and $m' = 1$ if $m = 2$. The rightmost equality in eq. (S.14) follows immediately from eqs. (S.5)-(S.6). So one has a formalism where both the voltages and the dissipated heats are described as stochastic processes, driven by the thermal noises $\eta_m$.

### III. PROBABILITY DISTRIBUTION FUNCTION FOR THE VOLTAGES

We now study the joint probability distribution function (PDF) $P(V_1, V_2, t)$, that the system at time $t$ has a voltage drop $V_1$ across the resistor $R_1$ and a voltage drop $V_2$ across the resistor $R_2$. As the time evolution of $V_1$ and $V_2$ is described by the Langevin equations (S.7)-(S.8), it can be proved that the time evolution of $P(V_1, V_2, t)$ is governed by the Fokker-Planck equation [2]

$$\partial_t P(V_1, V_2, Q_1, t) = L_0 P(V_1, V_2, t) = -\frac{\partial}{\partial V_1}(f_1 P) - \frac{\partial}{\partial V_2}(f_2 P) + \theta_{11}\frac{\partial^2}{\partial V_1^2}P + \theta_{22}\frac{\partial^2}{\partial V_2^2}P$$
$$+ 2\theta_{12}\frac{\partial^2}{\partial V_1 \partial V_2}P \tag{S.15}$$

We are interested in the long time steady state solution of eq. (S.15), which is time independent $P(V_1, V_2, t \to \infty) = P_{ss}(V_1, V_2)$. As the deterministic forces in eqs. (S.7)-(S.8) are linear in the variables $V_1$ and $V_2$, such a steady state solution reads

$$P_{ss}(V_1, V_2) = \frac{2\pi e^{-(aV_1^2 + bV_1V_2 + cV_2^2)}}{\sqrt{-b^2 + 4ca}} \tag{S.16}$$

where the coefficients

$$a = \frac{X\{C_1T_2Y + C[CR_2T_1 + T_2(C_1R_1 + CR_1 + C_2R_2)]\}}{2[Y^2T_1T_2 + C^2R_1R_2(T_1 - T_2)^2]},$$
$$b = -\frac{XC[(C_2 + C)R_2T_1 + (C_1 + C)R_1T_2]}{[Y^2T_1T_2 + C^2R_1R_2(T_1 - T_2)^2]},$$
$$c = \frac{X\{C_2T_1Y + C[CR_1T_2 + T_1(C_1R_1 + CR_2 + C_2R_2)]\}}{2[Y^2T_1T_2 + C^2R_1R_2(T_1 - T_2)^2]},$$

can be obtained by replacing eq. (S.16) into eq. (S.15), and by imposing the steady state condition $\partial_t P = 0$. We are furthermore interested in the unconstrained steady state probabilities $P_{1,ss}(V_1)$, and $P_{2,ss}(V_2)$, which are obtained as follows

$$P_{1,ss}(V_1) = \int dV_2 P_{ss}(V_1, V_2) = \frac{e^{-\frac{V_1^2}{2\sigma_1^2}}}{\sqrt{2\pi\sigma_1^2}} \tag{S.17}$$

$$P_{2,ss}(V_2) = \int dV_1 P_{ss}(V_1, V_2) = \frac{e^{-\frac{V_2^2}{2\sigma_2^2}}}{\sqrt{2\pi\sigma_2^2}} \tag{S.18}$$

where the variances read

$$\sigma_1^2 = \frac{T_1(C + C_2)Y + (T_2 - T_1)C^2 R_1}{XY} \tag{S.19}$$

$$\sigma_2^2 = \frac{T_2(C + C_1)Y - (T_2 - T_1)C^2 R_2}{XY} \tag{S.20}$$

### IV. PROBABILITY DISTRIBUTION FOR THE DISSIPATED HEAT AND AVERAGE RATE

We start by noticing that the heat injected from the bath 1 is then dissipated in the bath 2 (and vice-versa), and so we expect the probability distribution of $Q_1$ and $Q_2$ to be symmetric. Thus in the following, we will only study the probability distribution of $Q_1$. We now proceed by introducing the joint probability distribution function of the variables $V_1$, $V_2$, and $Q_1$, $\Phi(V_1, V_2, Q_1, t)$ As each of these three variables evolves according to a Langevin equation, the time evolution of their PDF is described by the Fokker-Planck equation [3, 4]

$$\partial_t \Phi(V_1, V_2, Q_1, t) = -\frac{\partial}{\partial V_1}(f_1\Phi) - \frac{\partial}{\partial V_2}(f_2\Phi) + \theta_{11}\frac{\partial^2}{\partial V_1^2}\Phi + \theta_{22}\frac{\partial^2}{\partial V_2^2}\Phi + 2\theta_{12}\frac{\partial^2}{\partial V_1 \partial V_2}\Phi$$
$$-\frac{\partial}{\partial Q_1}\left\{r_{11}\left[\frac{\partial}{\partial V_1}(V_1\Phi) + \left(V_1\frac{\partial}{\partial V_1}\Phi\right)\right] + 2r_{12}\frac{\partial}{\partial V_2}(V_1\Phi) + \frac{V_1^2}{R_1}\Phi\right\}$$
$$+ V_1^2 r_{22}\frac{\partial^2}{\partial Q_1^2}\Phi \tag{S.21}$$

with

$$\begin{aligned} r_{11} &= k_1\theta_{11} + k_2\theta_{12}, \\ r_{12} &= k_1\theta_{12} + k_2\theta_{22}, \\ r_{22} &= k_1^2\theta_{11} + k_2^2\theta_{22} + 2k_1k_2\theta_{12}, \end{aligned} \quad (S.22)$$

and $k_1 = (C_1 + C)$, $k_2 = -C$. It is worth noting that the first part of the right hand side of eq. (S.21) is identical to the rhs of eq. (S.15).

We proceed by proving eq. (7) in the main text, expressing the dissipated heat rate as a function of the system parameters. We have

$$\begin{aligned} \partial_t \langle Q_1 \rangle_t &= \partial_t \int dV_1 dV_2 dQ_1 \, Q_1 \Phi(V_1, V_2, Q_1, t) = \int dV_1 dV_2 dQ_1 \, Q_1 \partial_t \Phi(V_1, V_2, Q, t) = -r_{11} + \frac{1}{R_1}\langle V_1^2 \rangle \\ &= \frac{C^2 \Delta T}{XY}, \end{aligned} \quad (S.23)$$

where we have replaced the time derivative $\partial_t \Phi(V_1, V_2, Q, t)$ with the rhs of eq. (S.21) and used the equality $\langle V_1^2 \rangle = \sigma_1^2$, with $\sigma_1$ as given by eq. (S.19). This equation corresponds to the one given in the main text.

We can now obtain the expressions for the variance of $V_1$ and $V_2$, as introduced in the main text. Using eq. (S.23) we can express eq. (S.19) and eq. (S.20) in the following way:

$$\sigma_m^2 = \sigma_{m,\text{eq}}^2 + <\dot{Q}_m> R_m \quad (S.24)$$

where $\sigma_{m,\text{eq}}^2 = \frac{T_m(C+C_{m'})}{X}$ is the equilibrium value of $\sigma_m^2$ at $<\dot{Q}_m>= 0$.

## V. CONSERVATION LAW

We now turn our attention to eq. (2), in the main text, and provide a formal proof for it. In order to do this, we derive a relation between the reservoir entropy change $\Delta S_{r,\tau}$ and the system dynamics. For simplicity, in the following we divide the time into small intervals $\Delta t$: let us assume that the system (the circuit in our case) is in the state $\mathbf{V} = (V_1, V_2)$ at time $t$, and let's denote by $\mathbf{V}' = (V_1 + \Delta V_1, V_2 + \Delta V_2)$ its state at time $t + \Delta t$. Let $\mathcal{P}_F(\mathbf{V} \to \mathbf{V}'|\mathbf{V}, t)$ be the probability that the system undergoes a transition from $\mathbf{V}$ to $\mathbf{V}'$ provided that its state at time $t$ is $\mathbf{V}$, and let $\mathcal{P}_R(\mathbf{V}' \to \mathbf{V}|\mathbf{V}', t + \Delta t)$ be the probability of the time-reverse transition. We have

$$\begin{aligned} \mathcal{P}_F(\mathbf{V} \to \mathbf{V}'|\mathbf{V}, t) &= \int d\eta_1 d\eta_2 \, \delta(\Delta V_1 - \Delta t \cdot (f1(V_1, V_2) + \sigma_{11}\eta_1 + \sigma_{12}\eta_2)) \\ &\times \delta(\Delta V_2 - \Delta t \cdot (f_2(V_1, V_2) + \sigma_{21}\eta_1 + \sigma_{22}\eta_2))p_1(\eta_1)p_2(\eta_2), \end{aligned} \quad (S.25)$$

where $\delta(x)$ is the Dirac delta function. Given that the noises are Gaussian distributed, their probability distributions read

$$p_m(\eta_m) = \exp\left[-\frac{\eta_m^2 \Delta t}{4R_m k_B T}\right] \sqrt{\frac{\Delta t}{4\pi R_m k_B T_m}} \quad (S.26)$$

and expressing the Dirac delta in Fourier space $\delta(x) = 1/(2\pi) \int dq \exp(iqx)$, eq. (S.25) becomes

$$\begin{aligned} \mathcal{P}_F(\mathbf{V} \to \mathbf{V}'|\mathbf{V}, t) &= \frac{1}{(2\pi)^2} \int dq_1 dq_2 \exp[\imath(q_1\Delta V_1 + q_2\Delta V_2)] \int \prod_m d\eta_m \, e^{\Delta t\left[\imath q_m(f_m + \sigma_{m1}\eta_1 + \sigma_{m2}\eta_2) - \frac{\eta_m^2}{4R_m k_B T}\right]} \quad (S.27) \\ &= \exp\left\{-\frac{\Delta t}{4k_B T_1 T_2}\left[C_1^2 R_1 T_2(\dot{V}_1 - f_1)^2 + C_2^2 R_2 T_1(\dot{V}_2 - f_2)^2 \right.\right. \\ &\quad + 2C(\dot{V}_1 - f_1 - \dot{V}_2 + f_2)(C_1 R_1 T_2(\dot{V}_1 - f_1) - C_2 R_2 T_1(\dot{V}_2 - f_2)) \\ &\quad \left.\left. + C^2(R_2 T_1 + R_1 T_2)(\dot{V}_1 - f_1 - \dot{V}_2 + f_2)^2\right]\right\} \frac{X}{4\pi k_B \Delta t}\sqrt{\frac{R_1 R_2}{T_1 T_2}}; \end{aligned} \quad (S.28)$$

where we have taken $\Delta V_m/\Delta t \simeq \dot{V}_m$, and exploited the fact that all the integrals in eq. (S.27) are Gaussian integrals. Similarly, for the reverse transition we obtain

$$\begin{aligned}
P_R(\mathbf{V}' \to \mathbf{V}|\mathbf{V}', t+\Delta t) &= \int d\eta_1 d\eta_2\, \delta(\Delta V_1 + \Delta t(f1(V_1', V_2') + \sigma_{11}\eta_1 + \sigma_{12}\eta_2)) \\
&\quad \times \delta(\Delta V_2 + \Delta t(f_2(V_1', V_2') + \sigma_{21}\eta_1 + \sigma_{22}\eta_2))p_1(\eta_1)p_2(\eta_2) \quad (S.29)\\
&= \exp\left\{-\frac{\Delta t}{4k_B T_1 T_2}\left[C_1^2 R_1 T_2(\dot{V}_1+f_1)^2 + C_2^2 R_2 T_1(\dot{V}_2+f_2)^2\right.\right.\\
&\quad +2C(\dot{V}_1+f_1-\dot{V}_2-f_2)(C_1 R_1 T_2(\dot{V}_1+f_1)-C_2 R_2 T_1(\dot{V}_2+f_2))\\
&\quad \left.\left.+C^2(R_2 T_1+R_1 T_2)(\dot{V}_1+f_1-\dot{V}_2-f_2)^2\right]\right\}\frac{X}{4\pi k_B \Delta t}\sqrt{\frac{R_1 R_2}{T_1 T_2}}. \quad (S.30)
\end{aligned}$$

We now consider the ratio between the probability of the forward and backward trajectories, and by substituting the explicit definitions of $f_1(V_1, V_2)$ and $f_2(V_1, V_2)$, as given by eqs. (S.9)-(S.10), into eqs. (S.28) and (S.30), we finally obtain

$$\log\frac{P_F(\mathbf{V}\to\mathbf{V}'|\mathbf{V},t)}{P_R(\mathbf{V}'\to\mathbf{V}|\mathbf{V}',t+\Delta t)} = -\Delta t\left(V_1\frac{(C_1+C)\dot{V}_1-C\dot{V}_2}{k_B T_1}+V_2\frac{(C_2+C)\dot{V}_2-C\dot{V}_1}{k_B T_2}\right)=\Delta t\left(\frac{\dot{Q}_1}{k_B T_1}+\frac{\dot{Q}_2}{k_B T_2}\right), \quad (S.31)$$

where we have exploited eq. (S.14) in order to obtain the rightmost equality. Thus, by taking a trajectory $\mathbf{V}\to\mathbf{V}'$ over an arbitrary time interval $[t, t+\tau]$, and by integrating the right hand side of eq. (S.31) over such time interval, we finally obtain

$$k_B \log\frac{\mathcal{P}_F(\mathbf{V}\to\mathbf{V}'|\mathbf{V},t)}{\mathcal{P}_R(\mathbf{V}'\to\mathbf{V}|\mathbf{V}',t+\tau)} = \left(\frac{Q_1}{T_1}+\frac{Q_2}{T_2}\right) = \Delta S_{r,\tau} \quad (S.32)$$

We now note that the system is in an out-of-equilibrium steady state characterized by a PDF $P_{ss}(V_1, V_2)$, and so, along any trajectory connecting two points in the phase space $\mathbf{V}$ and $\mathbf{V}'$ the following equality holds

$$\begin{aligned}
\exp[\Delta S_{tot}/k_B] &= \exp\left[(\Delta S_{r,\tau}+\Delta S_{s,\tau})/k_B\right]\\
&= \frac{\mathcal{P}_F(\mathbf{V}\to\mathbf{V}'|\mathbf{V},t)P_{ss}(\mathbf{V})}{\mathcal{P}_R(\mathbf{V}'\to\mathbf{V}|\mathbf{V}',t+\tau)P_{ss}(\mathbf{V}')}, \quad (S.33)
\end{aligned}$$

where we have exploited eq. (S.32), and the definition of $\Delta S_{s,\tau}$ in eq. (1) in the main text. Thus we finally obtain

$$\mathcal{P}_F(\mathbf{V}\to\mathbf{V}'|\mathbf{V},t)P_{ss}(\mathbf{V})\exp[-\Delta S_{tot}/k_B] = \mathcal{P}_R(\mathbf{V}'\to\mathbf{V}|\mathbf{V}',t+\tau)P_{ss}(\mathbf{V}') \quad (S.34)$$

and summing up both sides over all the possible trajectories connecting any two points $\mathbf{V}, \mathbf{V}'$ in the phase space, and exploiting the normalization condition of the backward probability, namely

$$\sum_{\mathbf{V}',\mathbf{V}} \mathcal{P}_R(\mathbf{V}'\to\mathbf{V}|\mathbf{V}',t+\tau)P_{ss}(\mathbf{V}') = 1, \quad (S.35)$$

one obtains eq. (2). It is worth noting that the explicit knowledge of $P_{ss}(\mathbf{V})$ is not required, in order to prove eq. (2).

Finally, we note that, from a general perspective, eqs. (S.7)-(S.8) correspond to the Langevin equations of a stochastic system, whose variables $V_1$ and $V_2$ interact through non-conservative forces, and where the white noise is correlated. Therefore our proof of eq. (S.35), and thus of eq. (2) in the main text, holds in general for systems with such characteristics.

---

[1] G. Cannatá, G. Scandurra, C. Ciofin, Rev. Scie. Instrum. 80, 114702 (2009).
[2] R. Zwanzig, Nonequilibrium Statistical Mechanics, Oxford University Press, Oxford, 2001.
[3] A. Imparato, L. Peliti, G. Pesce, G. Rusciano, A. Sasso, *Phys. Rev. E*, **76**: 050101R (2007).
[4] H. C. Fogedby, A. Imparato, *J. Stat. Mech.* P04005 (2012).




\begin{thebibliography}{0}
\expandafter\ifx\csname natexlab\endcsname\relax\def\natexlab#1{#1}\fi
\expandafter\ifx\csname bibnamefont\endcsname\relax
  \def\bibnamefont#1{#1}\fi
\expandafter\ifx\csname bibfnamefont\endcsname\relax
  \def\bibfnamefont#1{#1}\fi
\expandafter\ifx\csname citenamefont\endcsname\relax
  \def\citenamefont#1{#1}\fi
\expandafter\ifx\csname url\endcsname\relax
  \def\url#1{\texttt{#1}}\fi
\expandafter\ifx\csname urlprefix\endcsname\relax\def\urlprefix{URL }\fi
\providecommand{\bibinfo}[2]{#2}
\providecommand{\eprint}[2][]{\url{#2}}

\end{thebibliography}


\begin{thebibliography}{widest-label}

\bibitem{Seifert_2012} U. Seifert
\emph{Rep.  Progr. Phys.} {\bf 75}, 126001, (2012).
\bibitem{sek10} Sekimoto, K. \emph{Stochastic Energetics}. (Springer, 2010).
\bibitem{bli06} Blickle V., Speck T., Helden L.,  Seifert U., and Bechinger C.  
 \emph{Phys. Rev. Lett.} {\bf 96}, 070603 (2006).
\bibitem{Jop} Jop, P., Petrosyan A. and  Ciliberto S. 
\emph{EPL} {\bf 81}, 50005 (2008).
\bibitem{Ruben}  Gomez-Solano, J.~R.,   Petrosyan, A.,  Ciliberto, S.  Chetrite, R. and Gawedzki K. 
\emph{Phys. Rev. Lett} {103}, 040601 (2009).
\bibitem{Evans02}  G. M. Wang, E. M. Sevick, E. Mittag, D.  J. Searles, and D. J. Evans, {\it Phys. Rev. Lett.}, {\bf 89}: 050601 (2002).
\bibitem{seifbech2012} J. Mehl, B. Lander, C. Bechinger, B. Blicke and U. Seifert, 
 Phys. Rev. Lett. \textbf{108}, 220601 (2012).
\bibitem{kumiko}
K. Hayashi, H. Ueno, R. Iino,  H.  Noji, Phys. Rev. Lett. 104, 218103 (2010)
\bibitem{Ciliberto} S Ciliberto, S Joubaud and A Petrosyan
J. Stat. Mech., P12003 (2010).
\bibitem{gallavotti}
D. J. Evans \emph{et al.}, Phys. Rev. Lett. \textbf{71}, 2401 (1993);
G. Gallavotti, E. G. D. Cohen, J. Stat. Phys.
\textbf{80}, 931 (1995).
\bibitem{Deridda} T. Bodinau, B. Deridda, 
 Phy.Rev. Lett 92, 180601 (2004).
\bibitem{Jarz2004}  {C. Jarzynski and D. K. W\'ojcik Phys. Rev. Lett. 92, 230602 (2004).}
\bibitem{VandenBroeck} C. Van den Broeck, R. Kawai and P. Meurs,
Phys. Rev. Lett 93, 090601 (2004). 
\bibitem{Visco} P Visco, 
J. Stat. Mech., page
P06006, (2006).
\bibitem{Dhar2007}{ K. Saito and A. Dhar Phys. Rev. Lett. 99, 180601 (2007).}
\bibitem{Gas2009} {D. Andrieux, P. Gaspard, T. Monnai, S. Tasaki, New J. Phys. 11, 043014 (2009).}
\bibitem{evans_temp}
{Evans D. , Searles D. J.  \and Williams S. R.}, 
{J. Chem. Phys.} {132},  {024501} {2010}.
\bibitem{Hanggi2011} { M. Campisi, P. Talkner, P. Hanggi, Rev. Mod. Phys.
83, 771 (2011)}
\bibitem{Villamania} A. Crisanti, A. Puglisi,
and D. Villamaina, 
 Phys. Rev. E 85, 061127 (2012)
\bibitem{Nyquist}  H. Nyquist, 
 Phys. Rev. 32, 110 (1928) 
\bibitem{Johnson} J. Johnson, 
Phys. Rev. 32, 97 (1928)
\bibitem{Feymann} R. P. Feynman, R. B. Leighton, and M. Sands, The
Feynman Lectures on Physics I (Addison-Wesley,
Reading, MA, 1963), Chap. 46.
\bibitem{Smoluchowski} M. v. Smoluchowski, Phys. Z. 13, 1069 (1912).
\bibitem{Abb2000}  {D. Abbott,  B. R. Davis, and J. M. R. Parrondo, AIP Conf. Proc. 511, 213 (2000).}
\bibitem{Sekimoto}  Sekimoto K, 
Prog. Theor. Phys. Suppl. 130, 17, (1998)
\bibitem{Supplementary} \albb{Supplementary Material}. 
\bibitem{Garnier}  N. Garnier, S. Ciliberto, 
Phys. Rev. E, 71, 060101(R) (2005)
\bibitem{Seifert} U. Seifert, 
  {\it Phys. Rev. Lett.}, {\bf 95}: 040602 (2005).
\bibitem{EVDB10}  M. Esposito, C. Van den Broeck, 
 {\it Phys. Rev. Lett.}, {\bf 104}, 090601 (2010).
\bibitem{VanZonCil} R. van Zon, S. Ciliberto, E. G. D. Cohen,
 {\it Phys. Rev. Lett.} {\bf 92}: 130601 (2004).
\bibitem{alb1} A. Imparato, L. Peliti, G. Pesce, G. Rusciano, A. Sasso,
{\it Phys. Rev. E}, {\bf 76}: 050101R (2007).
\bibitem{alb2}  H. C. Fogedby, A. Imparato, 
{\it J. Stat. Mech.}  P04005 (2012). 
\bibitem{RSI} 
G. Cannat\'a, G. Scandurra, C. Ciofin,
Rev. Scie. Instrum. 80,114702 (2009).
\bibitem{harada} 
{Harada T. \and Sasa S.-I.},{Phys. Rev. Lett.,}{95,}{130602}{(2005)}.



\end{thebibliography}
\end{document}